\documentclass[conference]{IEEEtran}
\IEEEoverridecommandlockouts
\usepackage{cite}
\usepackage{amsmath,amssymb,amsfonts}
\usepackage{algorithmic}
\usepackage{graphicx}
\usepackage{textcomp}
\usepackage{xcolor}
\usepackage{float}   
\usepackage{booktabs}  
\usepackage{caption} 
\usepackage{graphicx}     
\usepackage{float}        
\usepackage{subcaption}   
\def\BibTeX{{\rm B\kern-.05em{\sc i\kern-.025em b}\kern-.08em
    T\kern-.1667em\lower.7ex\hbox{E}\kern-.125emX}}
\begin{document}

\bstctlcite{IEEEexample:BSTcontrol}

\title{\LARGE \textit{RANGAN}: GAN-empowered Anomaly Detection in 5G Cloud RAN}

\author{
\IEEEauthorblockN{Douglas Liao$^1$, Jiping Luo$^1$, Jens Vevstad$^2$, Nikolaos Pappas$^1$}
\IEEEauthorblockA{$^1$ \textit{Department of Computer and Information Science, Link\"{o}ping University}, Campus Valla, Sweden \\
$^2$ Ericsson AB, Link\"{o}ping, Sweden \\
E-mail: douli502@student.liu.se, jiping.luo@liu.se, jens.vevstad@ericsson.com, nikolaos.pappas@liu.se}
\thanks{This work was supported in part by the Graduate School in Computer Science (CUGS) and by the ROBUST-6G Grant Agreement No. 101139068.}
}

\maketitle
\begin{abstract}
Radio Access Network (RAN) systems are inherently complex, requiring continuous monitoring to prevent performance degradation and ensure optimal user experience. The RAN leverages numerous key performance indicators (KPIs) to evaluate system performance, generating vast amounts of data each second. This immense data volume can make troubleshooting and accurate diagnosis of performance anomalies more difficult. Furthermore, the highly dynamic nature of RAN performance demands adaptive methodologies capable of capturing temporal dependencies to detect anomalies reliably.
In response to these challenges, we introduce \textbf{RANGAN}, an anomaly detection framework that integrates a Generative Adversarial Network (GAN) with a transformer architecture. To enhance the capability of capturing temporal dependencies within the data, RANGAN employs a sliding window approach during data preprocessing. We rigorously evaluated RANGAN using the publicly available RAN performance dataset from the Spotlight project \cite{sun-2024}. Experimental results demonstrate that RANGAN achieves promising detection accuracy, notably attaining an F1-score of up to $83\%$ in identifying network contention issues.

\end{abstract}


\section{Introduction}

Radio Access Network (RANs) is a key component of modern telecommunications infrastructure, enabling ubiquitous connectivity for mobile users across voice communications, mobile applications, and a wide range of online services \cite{ericssonRAN}. With the rapid growth in global mobile data consumption and the continual evolution of mobile network technologies, RANs have become increasingly complex, integrating a multitude of advanced technologies to meet rising user expectations for high-speed, reliable, and seamless connectivity.

To keep pace with these demands, communication service providers are under constant pressure to deliver enhanced network performance and user experience, while simultaneously managing energy efficiency and promoting sustainability \cite{ericsson2025building, ericsson2025sustainable}. However, the increasing sophistication and scale of RAN systems have introduced significant challenges in network monitoring, particularly in the timely detection of anomalies that may cause performance degradation or imminent failures.

Addressing these challenges is critical for maintaining network reliability, service quality, and delivering substantial economic value. Effective anomaly detection and mitigation contribute to reducing operational disruptions, improving customer satisfaction, and reinforcing the competitive position of network operators in an increasingly dynamic and demanding market landscape.

In the context of telecommunications, delivering higher throughput, lower latency, and more reliable service remains a central objective \cite{lajous-2023}. However, diagnosing and troubleshooting issues in large-scale systems such as RAN introduces a range of intricate challenges, including:
  \textbf{Scarcity of Labeled Data}: One of the primary obstacles in anomaly detection is the limited availability of labeled datasets for training and validation. In many real-world RAN scenarios, such labels are either unavailable or prohibitively costly to obtain \cite{qureshi2023toward}. 
  \textbf{Model Adaptability}: Maintaining the effectiveness of anomaly detection models over time is nontrivial, particularly in environments where traffic patterns and software configurations are subject to continuous change, as is common in RAN deployments~\cite{chen-2021, maimo-2018}.
  \textbf{Domain-Specific Model Performance}: Techniques that are successful in one domain often fail to generalize effectively to another. Anomaly detection models tend to be highly data-specific, and no single method consistently performs best across diverse RAN contexts \cite{isaac2025active}.
  \textbf{Data Quality}: Telecommunication data frequently contains noise and missing values, introducing uncertainty into the system. These issues complicate the task of distinguishing between normal behavior and true anomalies \cite{kim-2020}.

Traditionally, engineers have relied on rule-based systems and statistical techniques to detect anomalies in software systems \cite{denning-1987, lunt-1993, rouillard-2004, wong-2002}. While these approaches offer simplicity and interpretability, they depend heavily on predefined thresholds and heuristics. As a result, their effectiveness significantly diminishes in dynamic environments, such as modern RANs, where traffic patterns, system behaviors, and anomalies evolve continuously across multiple dimensions. A key advancement in this domain has been the integration of temporal and generative modeling techniques. Lin et al.~\cite{lin-2020} proposed a hybrid architecture combining a Variational Autoencoder (VAE) with a Long Short-Term Memory (LSTM) network, embedding the LSTM within the latent space of the VAE. This design leverages the VAE’s capacity for learning compact representations and the LSTM’s ability to model long-range temporal dependencies, yielding superior anomaly detection performance compared to standalone models.
To address the scalability and real-time processing demands of 5G networks, Maimó et al.~\cite{maimo-2018} developed a two-stage detection pipeline: a Deep Belief Network for rapid initial processing followed by an LSTM for temporal refinement. Similarly, Su et al.~\cite{sun-2024} proposed a lightweight edge-cloud architecture for anomaly detection in Open RAN environments. Their system incorporates a VAE for reducing data dimensionality at the edge, followed by a time-series imputation model and CausalNex for interpretable causal analysis in the cloud.
In parallel, Ashima et al.~\cite{ashima-2020} applied a sparse autoencoder and SHAP to detect and interpret anomalies in multivariate trace data from 4G LTE networks. Addressing spatial correlations in RAN data, Hasan et al.~\cite{hasan-2024} introduced \textit{Simba}, a framework that combines Graph Convolutional Networks (GCNs) with Transformer architectures to capture spatial-temporal dependencies for interference detection in mobile networks.


In this paper, we propose \textit{RANGAN}, an unsupervised anomaly detection framework designed to identify network contention in RAN. RANGAN integrates a Generative Adversarial Network (GAN) with a transformer-based architecture to effectively capture complex temporal dependencies inherent in time-series performance data. Additionally, we investigate the impact of varying sliding window sizes on detection performance, highlighting the trade-offs between temporal resolution and model effectiveness. The presented results demonstrate that our method achieves a promising detection accuracy, and attaining an F1-score of up to $83\%$ in identifying network contention issues.

\section{Methodology}

\subsection{Dataset and Preprocessing}

We utilize the SpotLight dataset \cite{sun-2024}, which simulates network traffic generated by five user equipments (UEs), each operating under distinct scenarios based on the traffic profiles summarized in Table~\ref{tab:traffic_profiles}. The system's key performance indicators (KPIs) are defined in accordance with the 3GPP standard, ensuring alignment with industry norms. These KPIs span a broad range of categories, as detailed in Table~\ref{tab:KPI}.

\begin{table}[H]
\centering
\caption{Different profiles of generated traffic.}
\label{tab:traffic_profiles}
\setlength{\tabcolsep}{5pt}
\begin{tabular}{ c c }
\toprule
\textbf{Traffic types} & \textbf{Description}  \\ 
\midrule
iperf3 TCP DL & TCP traffic from base station to UE \\ \hline
iperf3 TCP UL & TCP traffic from UE to base station \\ \hline
iperf3 UDP DL & UDP traffic at 10 Mbps from base station to UE  \\ \hline
iperf3 UDP UL & UDP traffic at 10 Mbps from UE to base station \\ \hline
file download & UE downloads a file from the Internet \\ \hline
file upload & UE uploads a file using scp \\ \hline
video stream & UE streams a video \\ \hline
web traffic & UE accesses random websites continuously \\ \hline
random ping & UE sends ping requests at varying rates and counts \\ 
\bottomrule
\end{tabular}

\vspace{1ex}
\footnotesize
\textbf{Notes:}  
scp = Secure Copy Protocol; DL = Downlink; UL = Uplink 

\end{table}

\vspace{-0.2in}

\begin{table}[H]
\centering
\caption{Summary of KPIs.}
\label{tab:KPI}
\begin{tabular}{c c }
\toprule
\textbf{KPI Category} & \textbf{Description}  \\ 
\midrule
Fronthaul Traffic  & UL/DL link usage (Gbps)  \\ \hline
Thread Scheduling & CPU runtime on/off  \\ \hline
PTP Logs & Frequency, RMS, delay, max offset  \\ 
\bottomrule
\end{tabular}

\vspace{1ex}
\footnotesize
\textbf{Notes:} PTP = Precision Time Protocol; RMS = Root Mean Square.

\end{table}

\begin{figure}[t]
\includegraphics[width=0.42\textwidth]{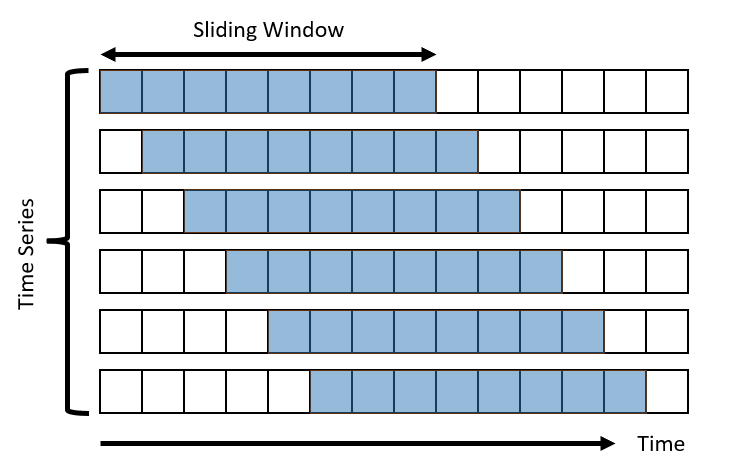}
\caption{Sliding window technique.}
\label{fig:window}
\end{figure}

The dataset was preprocessed by first selecting informative KPIs and then applying min-max normalization to scale the values within the range $[0, 1]$. Subsequently, a sliding window technique was employed to partition the time series into overlapping fixed-length segments, as illustrated in Figure~\ref{fig:window}. This approach enhances the model’s ability to learn and capture temporal patterns effectively.

\subsection{Proposed Model}

The anomaly detection architecture employed in this work is based on a GAN, augmented with transformer components. This design choice leverages the GAN’s proven capability to learn complex data distributions in an unsupervised setting. In combination with the sliding window technique, the model is particularly effective at identifying contextual anomalies in time-series data. 

\begin{figure}[ht]
\includegraphics[width=.45\textwidth]{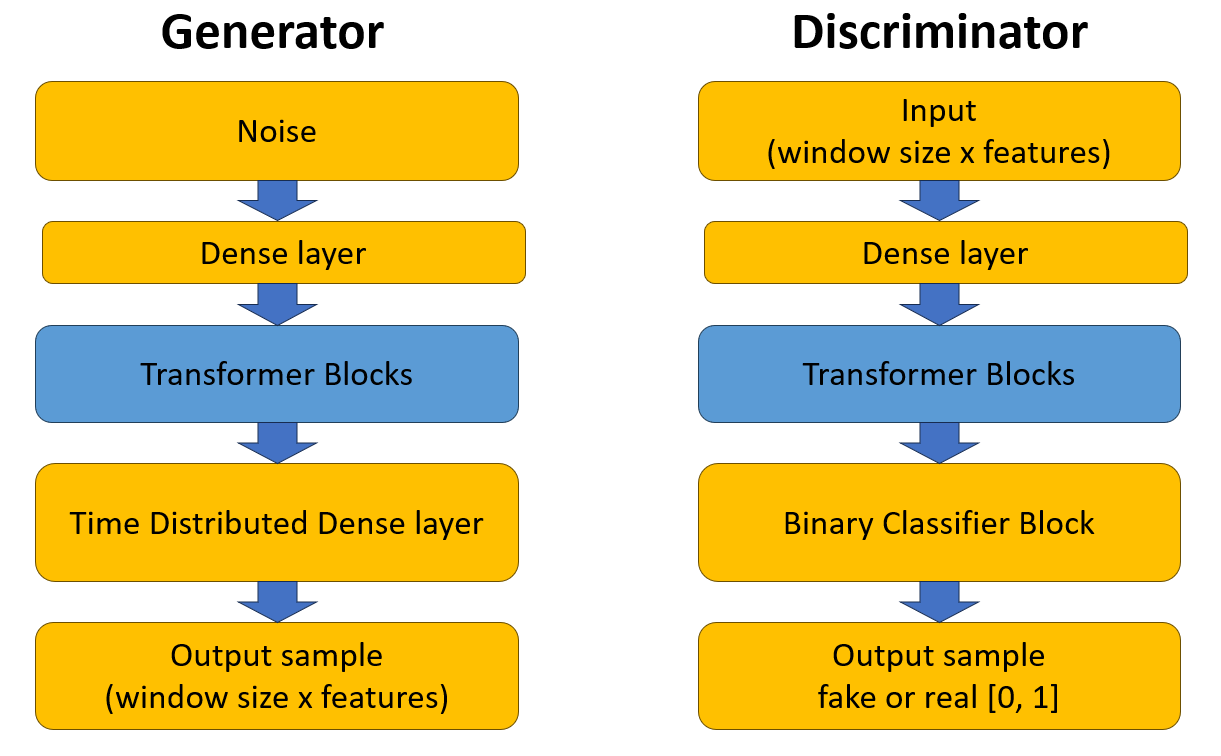}
\caption{The architecture of the proposed network.}
\label{fig:generator_discriminator}
\end{figure}

Figure~\ref{fig:generator_discriminator} illustrates the architectures of the generator and discriminator. The generator receives a latent input vector (noise) and produces outputs that match the dimensionality of the training samples. The discriminator, in turn, is trained to distinguish between real and synthetically generated samples, thereby guiding the generator to produce increasingly realistic outputs over time.


To enhance the model’s capability to capture temporal dependencies and complex structural patterns in the data, transformer blocks were integrated into both the generator and the discriminator. At the core of each transformer block lies the attention mechanism, which enables the model to represent contextual relationships more effectively. 

The attention mechanism operates by assigning dynamic weights to each time step based on its relevance to others in the sequence. This allows the model to selectively focus on the most informative segments of the input when generating or discriminating time-series data, thereby improving its overall performance in anomaly detection tasks.

\subsection{Evaluation Metrics}
We evaluate RANGAN using common metrics for anomaly detection.

\begin{equation}
\text{Precision} = \frac{TP}{TP + FP}
\end{equation}

\begin{equation}
\text{Recall} = \frac{TP}{TP + FN}
\end{equation}

\begin{equation}
F_1\text{-score} = 2 \times \frac{\text{Precision} \times \text{Recall}}{\text{Precision} + \text{Recall}}
\end{equation}

where $TP$ denotes true positives, $FP$ false positives, and $FN$ false negatives. The F1-score is particularly relevant in anomaly detection as it balances precision and recall, effectively capturing the trade-off between false alarms and missed anomalies. We also use the Area Under the Receiver Operating Characteristic Curve (ROC AUC) to assess the model’s ability to distinguish between normal and anomalous instances across various thresholds, limited to models producing continuous or probabilistic anomaly scores.

\section{Performance evaluation}


In this section, we assess the performance of RANGAN with different sliding window sizes on the dataset introduced in the Methodology. We then proceed to compare RANGAN’s results to several well-known anomaly detection methods: Autoencoder, Isolation Forest (IF), Local Outlier Factor (LOF), One-Class Support Vector Machine (OCSVM), Hierarchical Density-Based Spatial Clustering of Applications with Noise (HDBSCAN), Density-Based Spatial Clustering of Applications with Noise (DBSCAN), and ZScore.

As observed from Table~\ref{tab:network_performance}, among these methods, RANGAN achieved the best overall performance. The proposed model yielded a relatively high F1-score of $0.83$, a balanced precision of $0.75$, and a strong recall of $0.93$. In addition, RANGAN also demonstrated a solid ROC AUC value of $0.78$, suggesting that the model has a good ability to distinguish between normal and anomaly data, apart from the $1585$ FP. 

Furthermore, it can be observed that the autoencoder approach showed a competitive result, achieving an F1 score of $0.73$, a precision of $0.66$, and a recall of $0.83$. Likewise, as RANGAN, the autoencoder received a good ROC AUC value of $0.71$, indicating good discrimination capability. However, it yielded a much higher FP compared to RANGAN.

In contrast, most traditional methods, such as IF and LOF performed poorly in terms of the F1-score, $0.24$ and $0.20$ respectively. The remaining methods OCSVM, HDBSCAN, DBSCAN, and Z-score showed limited effectiveness, with a very low F1-score of $0.04$ each.

\begin{table}[ht]
\centering
\caption{Performance comparison of different methods.}
\label{tab:network_performance}
\renewcommand{\arraystretch}{0.5}
\begin{tabular}{lcccccc}
\toprule
\multicolumn{1}{c}{} & \multicolumn{5}{c}{\textbf{Network}} \\
\cmidrule(lr){2-6}
\textbf{Method} & \textbf{F1} & \textbf{Prec.} & \textbf{Rec.} & \textbf{ROC AUC} & \textbf{FP} \\
\midrule
RANGAN & 0.83 & 0.75 & 0.93 & 0.78 & 1585 \\
\addlinespace
Autoencoder  & 0.73 & 0.66 & 0.83 & 0.71 & 2080 \\
\addlinespace
IF  & 0.53 & 0.75 & 0.41 & 0.66 & 691 \\
\addlinespace
LOF & 0.20 & 0.66 & 0.12 & 0.56 & 310 \\
\addlinespace
OCSVM & 0.04 & 0.02 & 1.00 & 0.50 & 9092 \\
\addlinespace
HDBSCAN & 0.04 & 0.02 & 1.00 & NaN & 9092 \\
\addlinespace
DBSCAN  & 0.04 & 0.02 & 1.00 & NaN & 9092 \\
\addlinespace
ZScore & 0.04 & 0.02 & 0.85 & NaN & 6800 \\
\bottomrule
\end{tabular}
\end{table}

\begin{table}[ht]
\centering
\caption{The impact of the size of the sliding window.}
\label{tab:sliding_window}
\renewcommand{\arraystretch}{0.8}
\begin{tabular}{lcccccc}
\toprule
\multicolumn{1}{c}{} & \multicolumn{5}{c}{\textbf{Network}} \\
\cmidrule(lr){2-6}
\textbf{Method} & \textbf{F1} & \textbf{Prec.} & \textbf{Rec.} & \textbf{ROC AUC} & \textbf{FP} \\
\midrule
RANGAN/60 & 0.83 & 0.75 & 0.93 & 0.78 & 1585 \\
\addlinespace
RANGAN/50 & 0.83 & 0.74 & 0.94 & 0.78 & 1631 \\
\addlinespace
RANGAN/40 & 0.76 & 0.64 & 0.94 & 0.74 & 2253 \\
\addlinespace
RANGAN/30 & 0.63 & 0.48 & 0.93 & 0.69 & 3296 \\
\addlinespace
RANGAN/20 & 0.47 & 0.31 & 0.91 & 0.65 & 4311 \\
\bottomrule
\end{tabular}
\end{table}



The results presented in Table~\ref{tab:sliding_window} illustrate the impact of varying sliding window sizes on the performance of the proposed model. The evaluated window sizes range from $20$ to $60$, and a clear trend emerges: larger window sizes generally lead to improved performance. Notably, window sizes of $50$ and $60$ achieved the best results, both yielding an F1 score of $0.83$ and high recall values of $0.93$ and $0.94$, respectively. These configurations also exhibited similar precision and ROC AUC scores. The most noticeable difference between them lies in the number of FPs, with the window size of $60$ producing slightly fewer FPs than the size $50$ counterpart.

Although recall remains consistently high across all window sizes, a gradual decline in overall performance is observed as the window size decreases—particularly in terms of precision and the corresponding increase in false positives. \textit{This highlights the importance of sufficient temporal context in effectively distinguishing anomalous from normal behavior}.

\begin{figure}[ht]
    \centering
    \begin{subfigure}[b]{0.5\textwidth}
        \includegraphics[width=\textwidth]{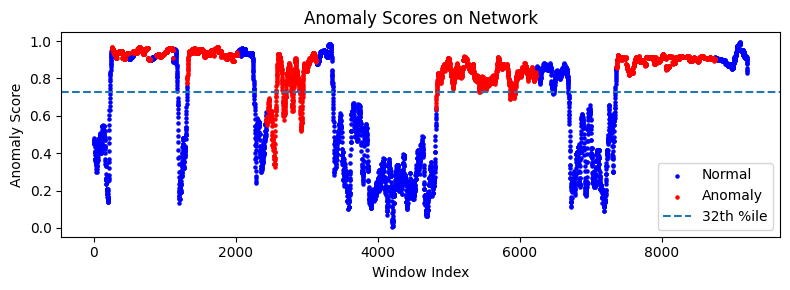}
        \caption{Window size = 50}
    \end{subfigure}
    \\
    \begin{subfigure}[b]{0.5\textwidth}
        \includegraphics[width=\textwidth]{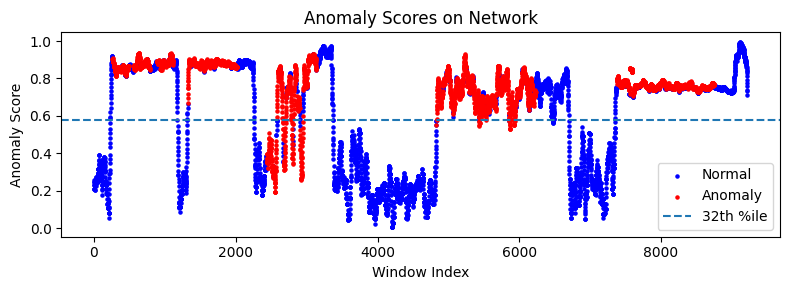}
        \caption{Window size = 40}
    \end{subfigure}
    \\
    \begin{subfigure}[b]{0.5\textwidth}
        \includegraphics[width=\textwidth]{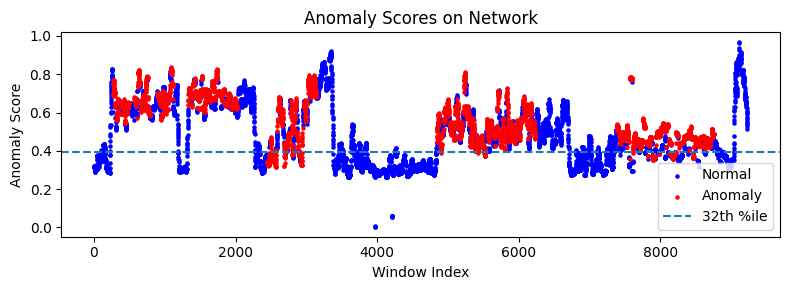}
        \caption{Window size = 30}
    \end{subfigure}
    \\
    \begin{subfigure}[b]{0.5\textwidth}
        \includegraphics[width=\textwidth]{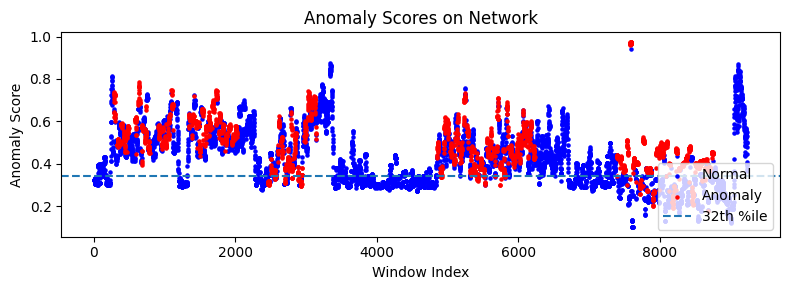}
        \caption{Window size = 20}
    \end{subfigure}
    \caption{Anomaly score of different window sizes.}
    \label{fig:anomaly_scores_stacked}
\end{figure}

Furthermore, Figure~\ref{fig:anomaly_scores_stacked} provides a visual comparison of the model’s anomaly scores across different sliding window sizes. These scores reflect the model’s confidence in the presence of an anomaly within each windowed segment, with higher values indicating stronger anomaly likelihood. As shown, \textit{larger window sizes yield a more distinct separation between normal and anomalous segments}, which helps explain the improved ROC AUC scores reported in Table~\ref{tab:sliding_window}. It is important to note that the number of detected anomalies in each plot varies with the window size, since a window is labeled anomalous if it contains any anomalous data point within its range.

\section{Conclusions and Future directions}

Our results demonstrate that deep learning models can achieve strong performance in detecting network contention within RAN. The proposed model, \textit{RANGAN}, attained the highest F1 score of $83\%$, while maintaining a moderate number of FP, many of which were observed to occur shortly after true anomalies. Although deep learning models exhibited slightly higher FP rates compared to traditional methods, this was offset by substantially higher recall and the ability to detect contextual anomalies that simpler techniques often failed to identify. Complementary techniques such as feature engineering and the application of sliding windows were shown to enhance detection performance, particularly for time-dependent datasets. Experimental results indicated that smaller window sizes introduced greater noise and uncertainty in detecting network contention. This trend was evident in both the anomaly score visualizations and the degradation in ROC AUC performance.

Despite the strong results achieved by RANGAN, several areas remain open for further research and refinement. A key limitation is the model’s current inability to generalize anomaly detection across different components of the system, such as PDCP thread contention, radio interference, and MAC thread contention. Future work should therefore consider expanding the detection scope, either through a unified model capable of handling all anomaly types or through specialized models trained for specific anomaly categories.

In practical deployments, anomaly detection represents only the initial step toward robust and reliable network operations. Once an anomaly is identified, it is crucial to determine its root cause in order to provide actionable insights before implementing corrective measures. Telecommunication stakeholders often prioritize causal understanding over mere anomaly detection. Consequently, future work should also explore the integration of explainability techniques as part of a comprehensive anomaly detection framework.

\bibliographystyle{IEEEtran}
\bibliography{references}

\end{document}